\documentclass{sig-alternate}
\pdfoutput=1

\def \imagedir {.}
\input{glyphtounicode}
\usepackage[utf8]{inputenc}
\usepackage[T1]{fontenc}

\usepackage[english]{babel}
\usepackage{graphicx}
\usepackage{caption}
\usepackage{subcaption}
\usepackage{balance}  				
\usepackage{listings}					
\usepackage{dblfloatfix}			

\newcommand{\ie}{i.\,e.}
\newcommand{\eg}{e.\,g.}

\newcommand\circa{{\raise.17ex\hbox{$\scriptstyle\mathtt{\sim}$}}}

\def\sharedaffiliation{%
\end{tabular}
\begin{tabular}{c}}

\begin{document}


\title{Energy and Performance---Can a \\Wimpy-Node Cluster Challenge a Brawny Server?}
\numberofauthors{2}
\author{
	\alignauthor
	Daniel Schall\\
  \email{schall@cs.uni-kl.de}
	\alignauthor
	Theo H\"arder\\
  \email{haerder@cs.uni-kl.de}
  \sharedaffiliation
   \affaddr{Databases and Information Systems Group}  \\
   \affaddr{University of Kaiserslautern, Germany}
}

\maketitle
\begin{abstract}
Traditional DBMS servers are usually over-provisioned for most of their daily workloads and, because they do not show good energy proportionality, waste a lot of energy while underutilized.
A cluster of small (wimpy) servers, where the number of nodes can dynamically adjust to the current workload, might offer better energy characteristics for these workloads.
Yet, clusters suffer from "friction losses" and may not be able to quickly adapt to the workload, whereas a single, brawny server delivers performance instantaneously.

In this paper, we compare a small cluster of lightweight nodes to a single server in terms of performance and energy efficiency.
We run several benchmarks, consisting of OLTP and OLAP queries at variable utilization to test the system's ability to adjust to the workloads.
To quantify possible energy saving and its conceivable drawback on query runtime, we evaluate our implementation on a cluster as well as on a single, brawny server and compare the results w.r.t. performance and energy consumption.
Our findings confirm that---based on the workload---energy can be saved without sacrificing too much performance.
\end{abstract}

\section{Introduction}
Saving energy is a concern in all areas of IT.
Studies have shown that single servers have potential for energy optimizations, but, in general, the best performing configuration is also the most energy efficient one \cite{DBLP:conf/sigmod/TsirogiannisHS10}.
This observation stems from the fact that the power spectrum between idle and full utilization of a single server is narrow and 50\% of its power is already consumed at idle utilization \cite{DBLP:journals/computer/BarrosoH07}.

Today's server hardware is not energy-proportional; at low utilization, hardware---mainly main memory and storage drives---consumes a significant amount of power.
Hence, about half of the maximum power of a server is already going to waste when idle.
Automatically scaling systems down when idle, thus preventing high idle power consumption of today’s servers is the main focus of energy proportionality.
Unfortunately, current hardware is not energy proportional.
Several components such as CPUs  are able to quickly change into sleep states, requiring less energy, when idle.
Other components, especially the two main energy consumers of DBMSs, main memory and external storage, exhibit bad energy characteristics.
\begin{figure*}[t]
	\centering
	\includegraphics[width=0.8\textwidth, page=3]{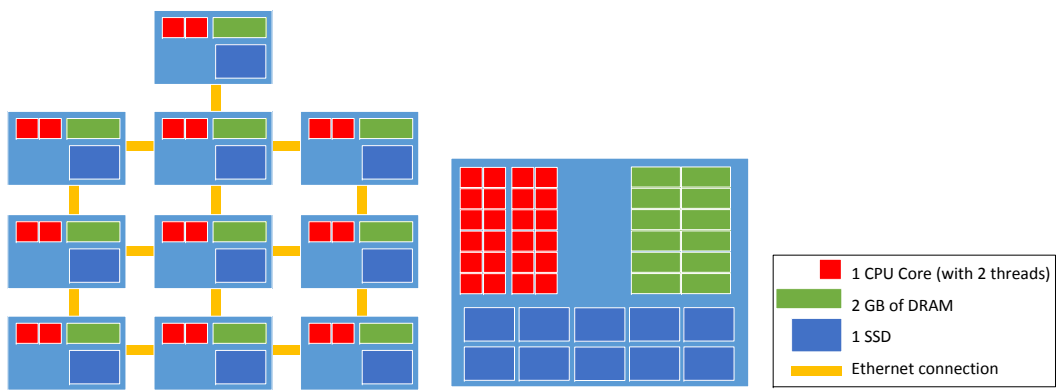}
	\vspace{0.3cm}
	\caption{The 10-node cluster compared with the brawny server}

	\label{figure:cluster}
\end{figure*}

Therefore, better energy efficiency cannot be achieved with current, centralized solutions.
This observation also holds for traditional DBMSs, composed of a single server with huge main memory and lots of storage drives attached.
In contrast to centralized, brawny servers, a scale-out cluster of lightweight (wimpy) servers has the ability to shutdown single nodes independently.
At an abstract level, this enables the cluster to dynamically add or remove storage and processing power based on the cluster's utilization.

With cloud computing, elastic systems have emerged that adapt their size to the current workload.
While stateless or lightweight systems can easily 
increase or reduce the number of active computing nodes in a cluster, a database faces much more challenges due to high interactions among the nodes and fast reachability of DB data.

Similar to cloud-based solutions, we hypothesize that a cluster of nodes may adjust the number of active (power-consuming) nodes to the current demand and, thus, approximate energy proportionality.

Based on these observations, we developed WattDB, a research prototype of a distributed DBMS cluster, running on lightweight, Amdahl-balanced nodes using commodity hardware.
The cluster is intended to dynamically shrink and expand its size, dependent on the workload.
Although the cluster may not be as powerful as a monolithic server, for typical workloads, we expect our system to consume significantly less energy.

Reconfiguring a cluster to dynamically match the workload requires data to be moved from node to node to balance utilization.
Yet, copying data is time-consuming and adds overhead to the already loaded cluster.
Reducing both, time and overhead, is crucial for an elastic DBMS.

In this paper, we compare a single, brawny server with a cluster of wimpy nodes under OLTP and OLAP workloads, running TPC-H and TPC-C respectively.
First, we give an overview of recent research addressing  partitioning, elasticity, and energy efficiency of DBMSs in Section~\ref{section:RelatedWork}.
In the following section, we introduce important aspects of our energy-proportional database cluster, called \emph{WattDB}.
Section~\ref{section:Experiments} contains the results of several empirical experiments and compares energy use and performance of our cluster to those of a brawny server.
In Section~\ref{section:Conclusion}, we summarize the main issues of our work and give some conclusions.

\section{Related Work}
\label{section:RelatedWork}
Reducing energy consumption of servers and dynamic reconfiguration are all subject to a variety of research approaches.
For the reason, we give a short overview of related works in three fields that serve as a building blocks of our research.

\subsection{Dynamic Clustering}
Traditional clustered DBMSs do not dynamically adjust their size (in terms of the number of active nodes) to their workload.
Hence, scale-out to additional nodes is typically supported, whereas the opposite functionality, shrinking the cluster and centralizing the processing---the so-called scale-in---, is not.
Recently, with the emergence of clouds,  a change of thinking occurred and dynamic solutions became a research topic.

In his PhD thesis \cite{Das:2011:SET:2521552}, Sudipto Das implemented an elastic data storage, called Elastras, able to dynamically grow and shrink on a cloud.
As common in generic clouds, his work is based on decoupled storage where all I/O involves network communication.
\emph{Key Groups}, an application-defined set of records frequently accessed together, 
can be seen as dynamic partitions that are often formed and dissolved.
By distributing the partitions among nodes in the cluster, both performance and cost can be controlled.

A lot more data management systems working on a cloud have been proposed.
In \cite{Brantner}, Brantner et al. designed a DBMS using Amazon S3 as storage and running on top.
Lomet et al. \cite{Lomet} divided the database into two layers, one transactional and one persistence component that can run independently.

In \cite{SCADS}, Armbrust et al. propose a scalable storage layer supporting consistency and dynamic scale-out/in called SCADS.
Objects in SCADS are stored in logical order. Hot, \ie, frequently accessed objects are distributed among disks to improve access latencies and mitigate bottlenecks.
The system was also extended to automatically adjust to workload changes and autonomously redistribute data.

Besides relational approaches, other implementations relax traditional DBMS properties to gain performance and simplify partitioning.
Yahoo PNUTS \cite{PNUTS}, Bigtable \cite{Bigtable}, and Cassandra\footnote{\url{http://cassandra.apache.org/}}, are example of systems sacrificing transaction or schema support and query power \cite{SCADS}.
Instead of arbitrary access patterns on the data, only primary key accesses to a single record are supported \cite{Vo:2010:TET:1920841.1920907}.

As an improvement, Amazon's SimpleDB\footnote{\url{http://aws.amazon.com/simpledb/}} allows transactions to access multiple records, but limits accesses to single tables.
Moreover, most current scalable data storage systems lack the rich data model of an RDBMS, which burdens application developers with data management tasks.
Yet, no \textit{fully-autonomous, clustered DBMS} exists which can provide ACID properties for transactions and SQL-like queries while dynamically adjusting its size to the current workload.

\subsection{Energy Optimizations}
\begin{figure*}%
	\centering
    \begin{subfigure}[b]{0.66\columnwidth}
  		\includegraphics[width=\textwidth,page=1]{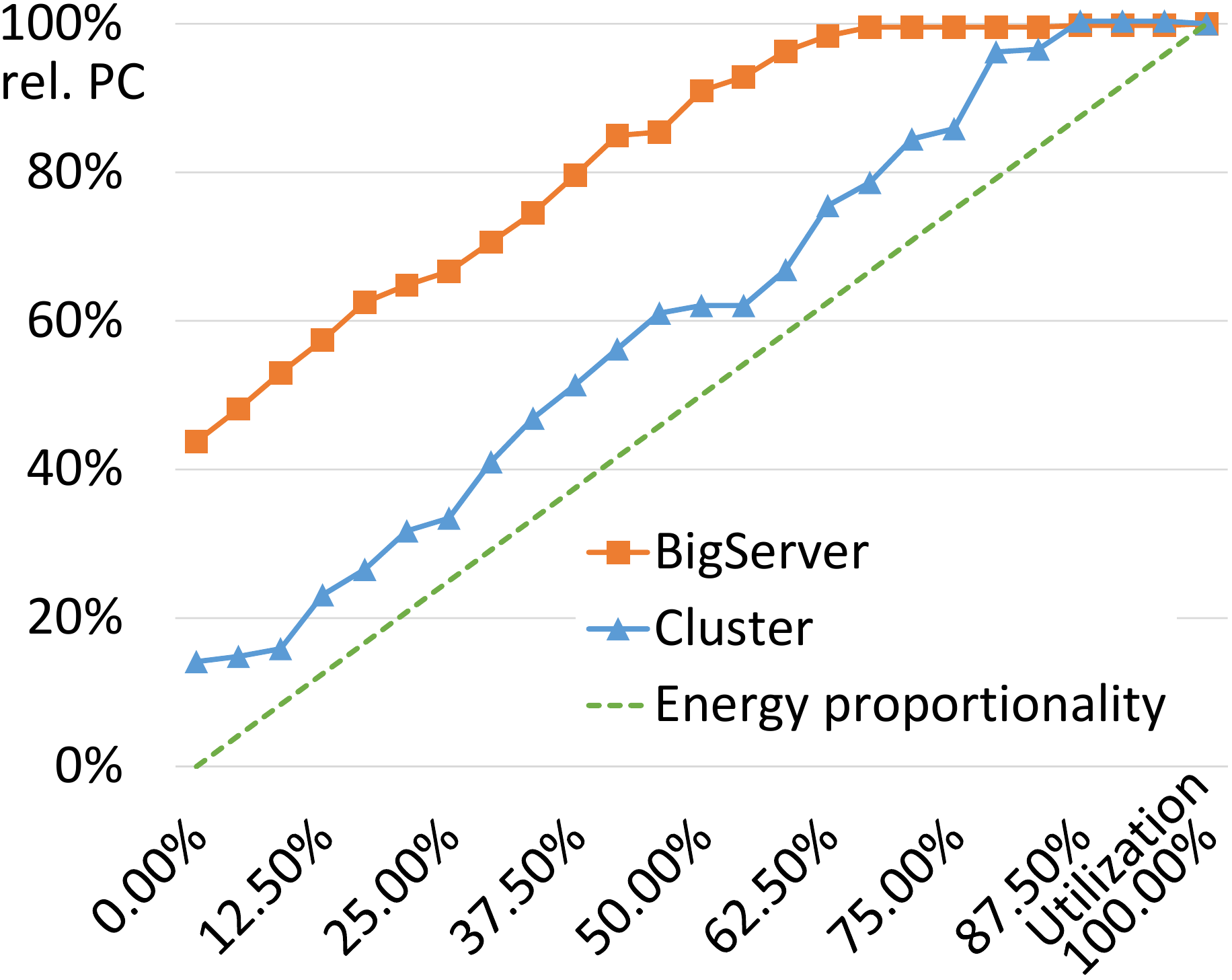}%
	  	\caption{Relative power consumption}%
		  \label{figure:pc:relative}%
	  \end{subfigure}%
		\hspace{0.02\columnwidth}
    \begin{subfigure}[b]{0.66\columnwidth}
  		\includegraphics[width=\textwidth,page=2]{\imagedir/PowerConsumption}%
	  	\caption{Power consumption}%
		  \label{figure:pc:absolute}%
	  \end{subfigure}%
		\hspace{0.02\columnwidth}
    \begin{subfigure}[b]{0.60\columnwidth}
		  \centering
  		\includegraphics[width=\textwidth,page=3]{\imagedir/PowerConsumption}%
	  	\caption{Theoretical performance}%
		  \label{figure:performance}%
	  \end{subfigure}%
		\caption{Power consumption and performance figures for both systems}
		\label{figure:clusterbigservercomparison}
\end{figure*}
Lang et al. \cite{Lang:2012:TED:2350229.2350280} have shown that a cluster suffers from ``friction losses'' due to coordination and data shipping overhead and is therefore not as powerful as a comparable heavyweight server.
On the other hand, for moderate workloads, \ie, the majority of real-world database applications, a scale-out cluster can exploit its ability to reduce or increase its size sufficiently fast and, in turn, gain far better energy efficiency.

In \cite{Schall:2013A}, we already explored the capabilities and limitations of a clustered storage architecture that dynamically adjusts the number of nodes to varying workloads consisting of simple \emph{read-only page requests} where a large file had to be accessed via an index\footnote{Starting our WattDB development and testing with rather simple workloads facilitated the understanding of the internal system behavior, the debugging process, as well as the identification of performance bottlenecks.}.
We concluded that it is possible to approximate energy proportionality in the storage layer with a cluster of wimpy nodes. However, attaching or detaching a storage server is rather expensive, because (parts of) datasets may have to be migrated. Therefore, such events (in appropriate workloads) should happen on a scale of minutes or hours, but not seconds.

In \cite{Schall:2013B}, we have focused on the query processing layer---again for varying workloads consisting of two types of \emph{read-only SQL queries}---and drawn similar conclusions.
In this contribution, we revealed that attaching or detaching a (pure) processing node is rather inexpensive, because repartitioning and movement of data is not needed.  Hence, such an event can happen in the range of a few seconds---without disturbing the current workload too much.

We substantially extended the kind of DBMS processing supported by WattDB to \emph{complex OLAP / OLTP workloads consisting of read-write transactions} in \cite{SH-DASFAA2014}.
For this purpose, we refined and combined both approaches to get one step closer to a fully-featured DBMS, able to process OLTP and OLAP workloads simultaneously.
In this work, we were able to trade performance for energy savings and vice versa.
Yet, we found out that the adaptation of the cluster and the data distribution to fit the query workload is time-consuming and needs to be optimized.

As discovered before, a single-server-based DBMS is far from being energy-proportional and cannot process realistic workloads in an energy-efficient way.
Our previous research indicates that a cluster of lightweight (wimpy) servers, where nodes can be dynamically switched on or off, seems more promising.
In this paper, compare our scale-out cluster to a big server to quantify possible energy savings and to discover promising workloads.

\section{Cluster vs. big Server}
Our cluster hardware consists of n (currently 10) identical nodes, interconnected by a Gigabit-ethernet switch.
Each node is equipped with an Intel Atom D510 CPU (with two threads using HyperThreading) running at 1.66 GHz, 2 GB of DRAM and an SSD for data storage. 
The configuration is considered Amdahl-balanced \cite{AmdahlBlade}, \ie, balanced w.r.t. I/O and network throughput on one hand and processing power on the other.
By choosing commodity hardware with limited data bandwidth, GB-Ethernet wiring is sufficient for interconnecting the nodes.
All nodes can communicate directly.

To compare performance and energy savings, we ran the same experiments again on a single, brawny server.
This server has two \emph{Intel Xeon X5670} processors with 24 GB of RAM and 10 SSDs.\footnote{For a fair comparison with the wimpy nodes, we have reduced the RAM to 24GB, although the server can handle much more. Yet, with more main memory, the power consumption of the server would also be much higher.}
Each CPU has 12 cores and 24 threads (using HyperThreading), running at 2.93 GHz.

Figure \ref{figure:cluster} sketches the cluster with 10 nodes and the big server.
For comparison, we have highlighted the main components (CPU cores, main memory and disk) inside the nodes as well as the communication network.

Each wimpy node consumes \circa22 -- 26 Watts when active (based on utilization) and \circa2.5 Watts in standby.	
The interconnecting network switch consumes 20 Watts and is included in all measurements.

In its minimal configuration---with only one node and the switch running and all other nodes in standby---the cluster consumes approx. 65 Watts.
This configuration does not include any disk drives, hence, a more realistic minimal configuration requires about 70 Watts.
In this state, a single node is serving the entire DBMS functionality (storage, processing, and cluster coordination).
With all nodes running at full utilization, the cluster will consume \circa260 to 280 Watts, depending on the number of disk drives installed.

This is another reason for choosing commodity hardware which uses much less energy compared to server-grade components.
For example, main memory consumes \circa2.5 Watts per DIMM module, whereas ECC memory, used in the brawny server, consumes \circa10 Watts per DIMM.

The power consumption of the brawny server (with 10 SSDs) ranges from \circa200 Watts when idle to \circa430 Watts at full utilization.\footnote{These measurements include only 24 GB of DRAM as previously explained.}
In theory, the systems should show similar performance.
All nodes in the cluster come with 16.6 (10x1.66) GFLOPS, whereas the performance of the big server is rated with 17.6 GFLOPS.
Furthermore, L2 caches and memory bandwidth of both systems are similar and the same number of disks is installed.
Figure \ref{figure:clusterbigservercomparison} depicts power consumption and performance figures, as given in the product sheets, for both systems.

\subsection{DBMS Software}
By the time, research gained interest in energy efficiency of database servers, no state-of-the-art DBMS was able to run on a dynamically adapting cluster.
To test our hypotheses (see Section 1), we developed 
\emph{WattDB} that supports SQL query processing with ACID properties, but is also able to adjust to the workload by scaling out or in, respectively.

To enable a fair comparison, the same software is running on the cluster and the big server.
On the latter, the dynamic features of WattDB are not needed and are therefore disabled.

The smallest configuration of WattDB is a single server, hosting all database functions and acting as endpoint to DB clients.
This server is called \emph{master node}.
DB objects (tab\-les, partitions) and query evaluation can be offloaded to arbitrary nodes in the cluster to relieve the node, but it will always act as the coordinator and client endpoint.

Some of the key features and design considerations of WattDB are explained in the following.

\subsection{Dynamic Query Processing}
To run queries on a cluster of nodes, distributed query plans are generated on the master node.
Except data access operators which need local access to the database's records, all query operators can be placed on remote nodes. 
Running query operators on a single node does not involve network communication among query operators, because all records are  transferred via main memory.
Distributing operators implies shipping of records among nodes and, hence, introduces network latencies.
Additionally, the bandwidth of the Gigabit Ethernet, which we are using for our experiments, is relatively small, compared to memory bandwidth.

To mitigate the negative effects of distribution, WattDB is using \emph{vectorized volcano-style query operators} \cite{Graefe:1994:VEP:627290.627558,conf/cidr/BonczZN05}; hence, operators ship a set of records on each call.
This reduces the number of calls between operators and, thus, network latencies.
To further decrease network latencies, buffering operators are used to prefetch records from remote nodes.
\emph{Buffering operators} act as  proxies between two (regular) operators; they asynchronously prefetch records, thus, hiding the delay of fetching the next set of records.

In WattDB, the query optimizer tries to put pipelining operators\footnote{Pipelining operators can process one record at a time and emit the result, \eg, projection operators.} on the same node to minimize latencies. Offloading pipeline operators to a remote node has little effect on workload balancing and, thus, does not pay off.
Instead, blocking operators\footnote{Blocking operators need to receive all records, before they can emit the first result record, \eg, sorting operators.} may be placed on remote nodes to equally distribute query processing. They generally consume more resources (CPU, main memory) and are therefore prime candidates for workload balancing in the cluster.
\begin{figure}%
	\centering
		\includegraphics[width=0.8\columnwidth,page=2]{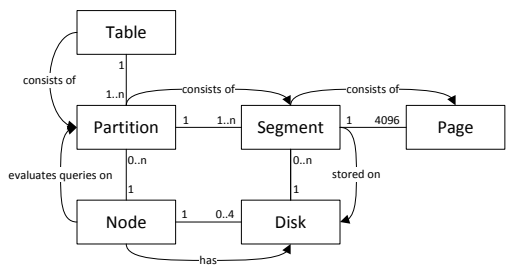}%
		\caption{Database schema}%
		\label{figure:schema}%
		\vspace{-0.3cm}
\end{figure}
\subsection{Dynamic Reorganization}
The master node is coordinating the whole cluster. It is globally optimizing the query plans, whereas regular nodes can locally optimize their part of the plan. Furthermore, it takes nodes on- and offline and decides when and how DB tables are (re)partitioned.

Every node is monitoring its 
utilization: CPU, memory consumption, network I/O, and disk utilization (storage and IOPS).
Additionally, 
performance-critical  data is collected for each database partition, \ie, CPU cycles, buffer page requests and network I/O.  With these figures, we can correlate the observed utilization of cluster components to  (logical) database entities.
Hence, both types of data are necessary to identify sources of 
cluster imbalance.
We use the performance figures of the components to identify their over- or under-utilization. In addition, activity recording 
of database entities is needed to determine the origin of the cluster's imbalance.
For this reason,
the nodes send their recording every few seconds to the master node.
\begin{figure*}[t]%
	\centering
    \begin{subfigure}[b]{0.9\columnwidth}
  		\includegraphics[width=\textwidth]{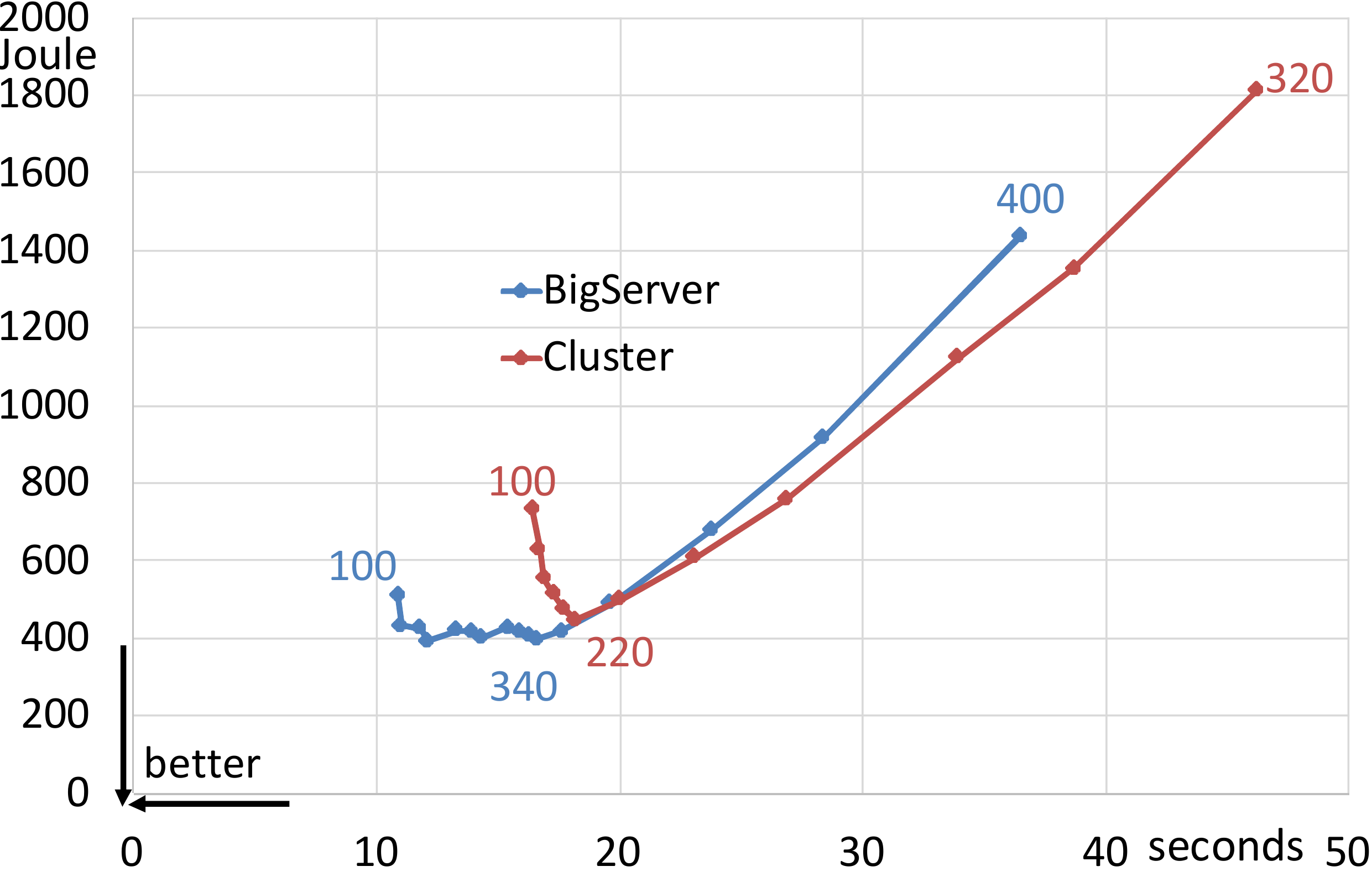}%
	  	\caption{OLAP: energy/query over response time}%
		  \label{figure:perf100:olap}%
	  \end{subfigure}%
		\hfill
    \begin{subfigure}[b]{0.9\columnwidth}
  		\includegraphics[width=\textwidth]{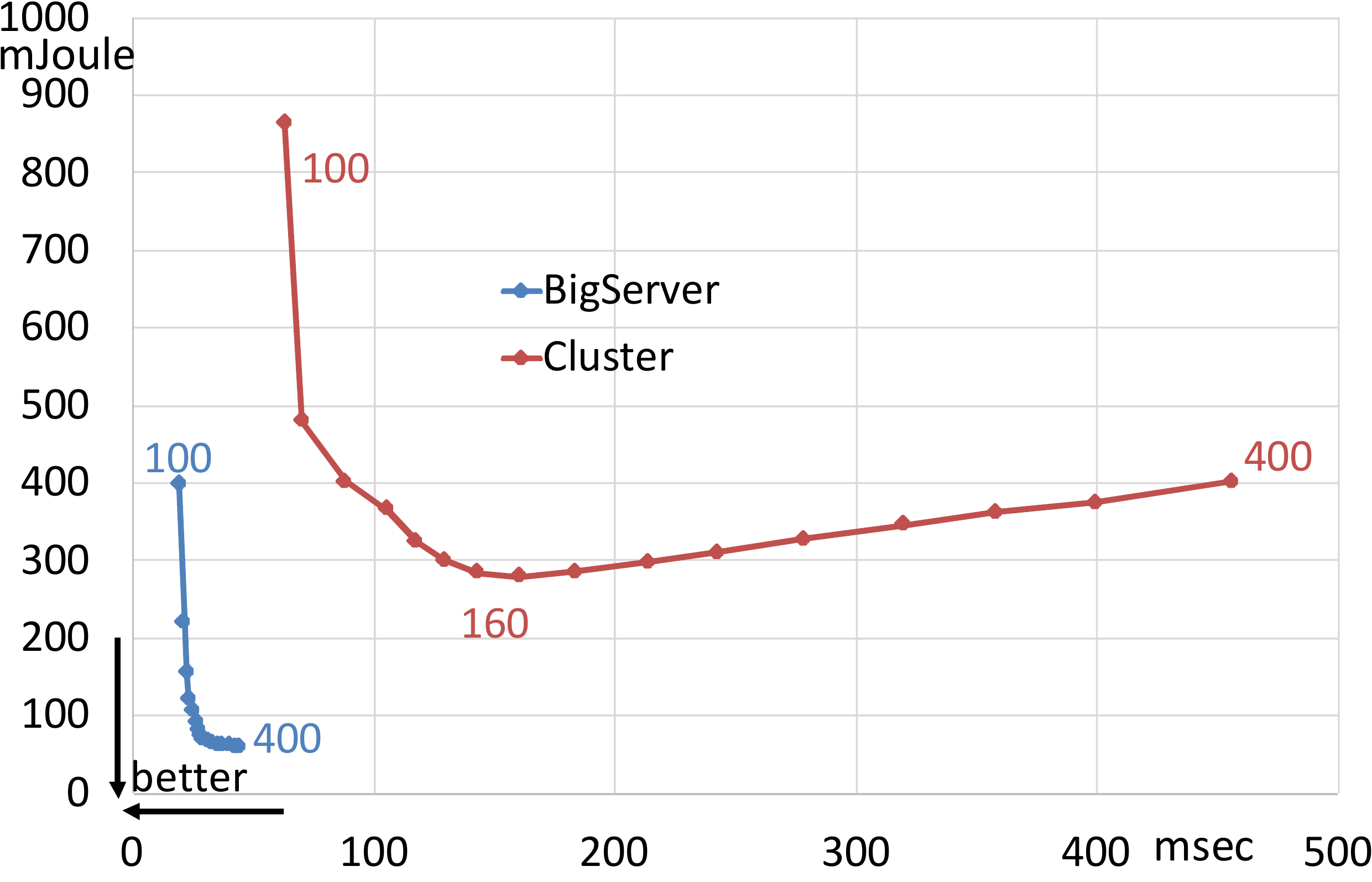}%
	  	\caption{OLTP: energy/query over response time}%
		  \label{figure:perf100:oltp}%
	  \end{subfigure}%
		\caption{Peak Performance and energy consumption for both systems}
		\label{figure:perf100}
\end{figure*}

The master checks the incoming performance data to predefined thresholds---with both upper and lower bounds.
If an overloaded component is detected, it will decide where to distribute data and whether to power on additional nodes and resume their cluster participation.
Similar, underutilized nodes trigger a scale-in protocol, \ie, the master will distribute the data (processing) to fewer nodes and shutdown the nodes currently not needed.
Decisions, what data to migrate and where, are done based on the current utilization of the nodes, the expected query workload, and the estimated cost, it will take to migrate data between nodes.

In WattDB, we have implemented different policies regarding the scale-out behavior.
First, each node in the cluster stores data on local disks to minimize network communication.
If storage space of a node is in short supply, database partitions are split up on nodes with free space.

Second, WattDB tries to keep the I/O rate for each storage disk in a certain range.
Underutilized disks are eligible for additional data---either newly generated by INSERT operations or migrated from overloaded disks.
Utilization among storage disks is first locally balanced on each node, before an allocation of data from/to other nodes is considered.

Third, each node's CPU utilization should not exceed the upper bound of the specified threshold (80\%).
As soon as this bound is violated for a node, WattDB first tries to off\-load query processing to underutilized nodes.\footnote{This works well for operators like \textbf{SORT}, \textbf{GROUP}, and \textbf{AGGREGATE}.} If the overload situation cannot be resolved by redistributing the query load, the current data partitions and their node assignments are reconsidered. When a partition causing the overload is identified, it is split according the partitioning scheme applied, where affected segments are moved to other nodes \cite{CIKM1}.
For underutilized nodes, an inverse approach is needed. A scale-in protocol is initiated, which quiesces the involved nodes from query processing and shifts their data partitions to nodes currently having sufficient processing capacity. 

Similar rules exist for network and memory utilization, \eg, if the working sets of the transactions become too big for the DB buffer, repartitioning is triggered.
WattDB makes decisions based on the current workload, the course of utilization in the recent past, and the expected future workloads \cite{KRAMER12}.
Additionally, workload shifts can be user-defined to inform the cluster of an expected change in utilization.

\textbf{Cost of reorganization}
Moving data is an expensive task, in terms of energy consumption and performance impact on concurrently running queries.
Data reorganization binds some computing resources, which would be needed to optimally process the query workload. This resource contention leads to fewer resources for the workload and, in turn, reduces query throughput.
However, the reorganization cost should amortize by reducing the energy consumption of subsequent queries.
Though it is difficult to calculate the exact energy consumption of a data move operation with respect to the impact of running queries, the energy cost can be estimated with the duration of the move operation and the (additional) power consumption.
Hence, moving 1 GByte of data to a dedicated node with 25 Watts power consumption will require approximately 10 seconds and 250 Joules.

In order to save energy, reconfiguration overhead needs to pay off by reducing query runtimes in the future.
Likewise, scale-in must trigger when the cluster is able to handle the workload with less nodes.
To estimate the impact of reorganization, WattDB relies on a simplified cost model where upcoming workload predictions and maintenance costs are calculated.

\subsection{Power Measurement}
We have developed a measurement device, capable of monitoring the power and energy consumption of each node in the cluster.
The device is also able to these metrics of the big server.
This device sends the stream of measurements to a connected PC, running the monitoring software.
The monitoring software can further capture the number of active database nodes and the total throughput and response times of queries during the tests.
This computer is controlling the benchmark execution by submitting queries to the master node; thus, it enables fine-grained monitoring in correlation with the benchmark runs.
The measurement frequency of the device reaches up to 100 Hz; hence, we are able to determine power use in high resolution.
A detailed description of the measurement device can be found in \cite{Schall:BTW2011}.

The energy measurement device is only used for external monitoring; WattDB cannot use the measurements to improve its energy efficiency.
Internally, the DBMS is working with estimates to determine overall power consumption.
\begin{figure*}[!t]%
	\centering
    \begin{subfigure}[b]{0.9\columnwidth}
  		\includegraphics[width=\textwidth,page=2]{\imagedir/CIKM_OLAP_0-100}%
	  	\caption{Query response time}%
		  \label{figure:perf0-100:olap:perf}%
	  \end{subfigure}%
		\hfill
    \begin{subfigure}[b]{0.9\columnwidth}
  		\includegraphics[width=\textwidth,page=1]{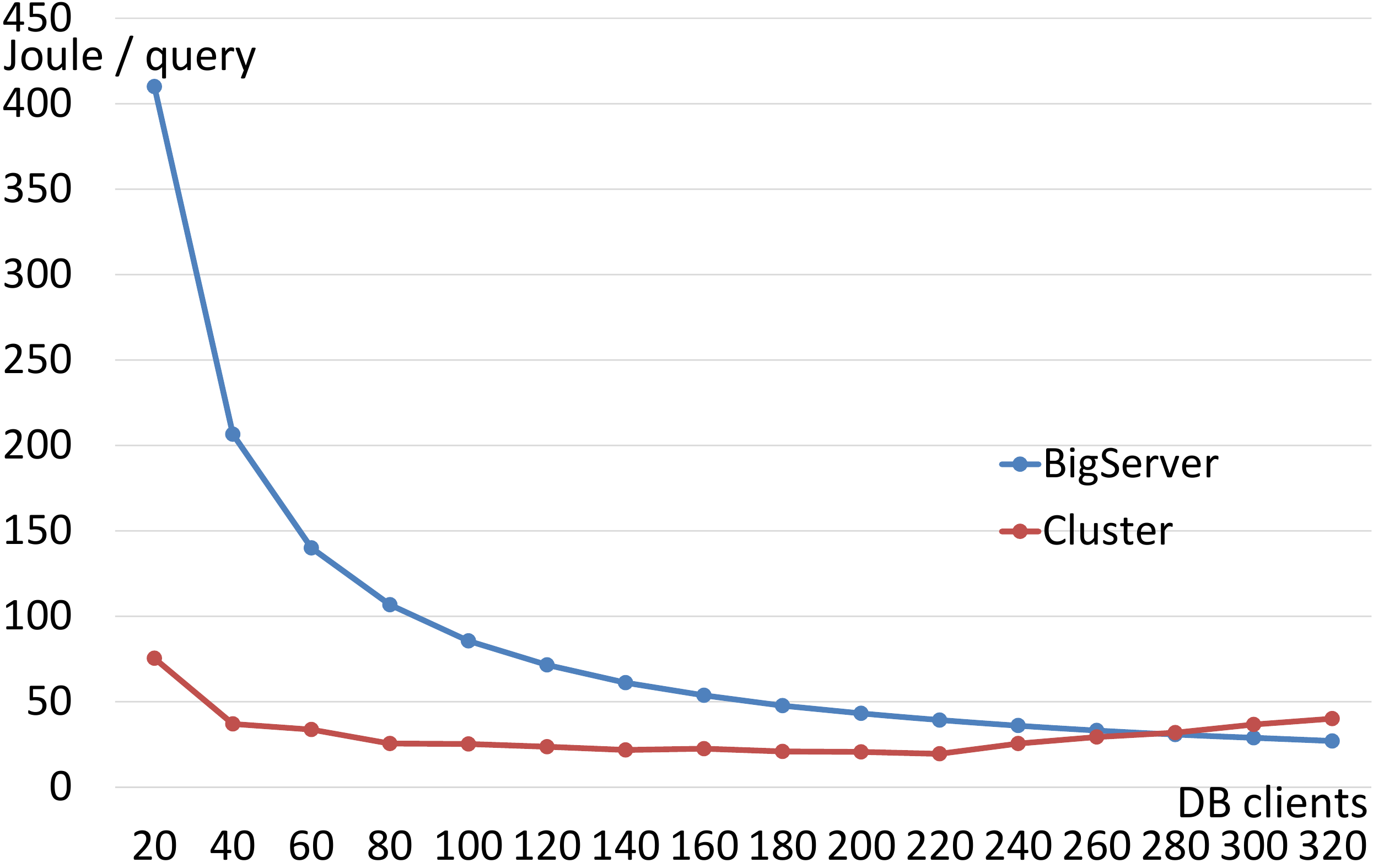}%
	  	\caption{Energy consumption per query}%
		  \label{figure:perf0-100:olap:ec}%
	  \end{subfigure}%
		\caption{Performance and energy consumption for varying OLAP utilizations}
		\label{figure:perf0-100:olap}
\end{figure*}

\begin{figure*}[!t]%
	\centering
    \begin{subfigure}[b]{0.9\columnwidth}
  		\includegraphics[width=\textwidth,page=2]{\imagedir/CIKM_OLTP_0-100}%
	  	\caption{Query response time}%
		  \label{figure:perf0-100:oltp:perf}%
	  \end{subfigure}%
		\hfill
    \begin{subfigure}[b]{0.9\columnwidth}
  		\includegraphics[width=\textwidth,page=1]{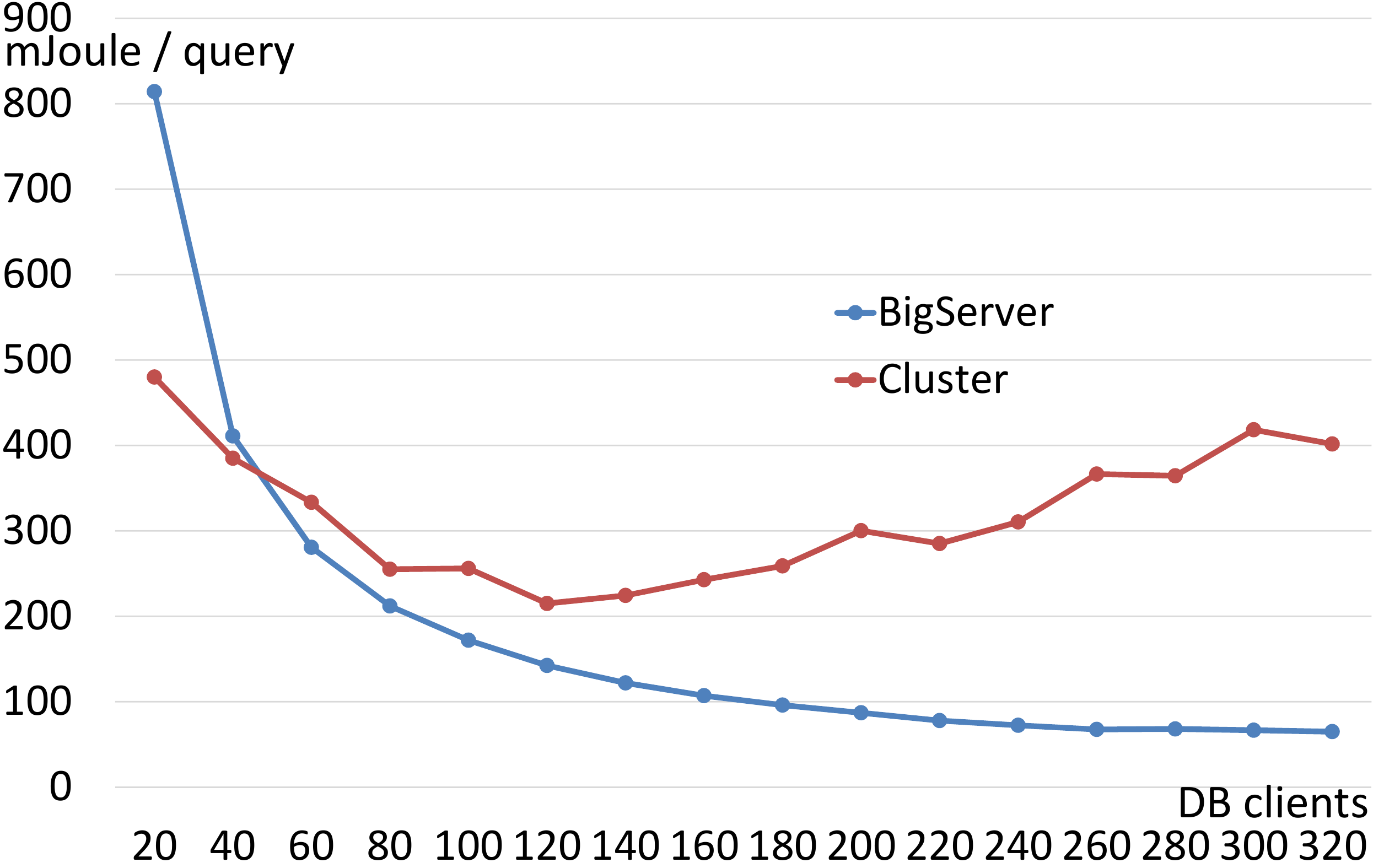}%
	  	\caption{Energy consumption per query}%
		  \label{figure:perf0-100:oltp:ec}%
	  \end{subfigure}%
		\caption{Performance and energy consumption for varying OLTP utilizations}
		\label{figure:perf0-100:oltp}
\end{figure*}

\section{Experiments}
\label{section:Experiments}
To compare the energy consumption of our cluster to that of a traditional DB server, we have processed OLAP and OLTP workloads on both platforms.
We have run performance-centric benchmarks first, to assess peak performance of both systems.
Next, we have evaluated energy-centric benchmarks to identify energy-efficiency potential of the big server and the cluster.
In the following, we first describe the experimental setup, before we present our results.

\subsection{Experimental Setup}
For all experiments, using OLTP and OLAP, we have set up the systems as previously described.
A separate server, directly connected to the master node and the big server, respectively,  is used as the benchmark driver, submitting queries to the cluster as well as monitoring response time and throughput.
The previously introduced power measurement device is also hooked up to the benchmark driver to correlate all measurements with energy consumption.

\textbf{OLAP workloads: }
For measuring OLAP performance and energy efficiency, we are using the well-known TPC-H benchmark with a scale factor of 300; hence, 300 GByte of raw data are generated.
Due to additional indexes and storage overhead, the final DB has approx. 460 GByte of raw data.
On the centralized server, small tables are stored on a single disk, whereas larger ones, \eg, the \textit{LINEITEM} and \textit{ORDERS} tables, are partitioned and distributed among all disks to increase access bandwidth and to parallelize processing on partitions.

On the cluster, the \textit{REGION} and \textit{NATION} tables are replicated to all nodes, while the other tables are partitioned and distributed equally among the nodes.
In static benchmarks, no repartitioning occurs, even if the initial distribution leads to hotspots in the data, that impact the node's performance.
If dynamic features of WattDB are enabled, the DBMS will automatically repartition as previously described.

\textbf{OLTP Workloads: }
For online transaction processing, we are running the TPC-C benchmark on the systems with a scale factor of 1000.
Hence, a thousand warehouses were generated on the cluster, consisting of about 100 GB of data.
Due to additional indexes and storage overhead, the final DB has approx. 200 GByte of raw data in the beginning of the experiments.

\subsection{Performance-centric benchmark}
First, to evaluate the \textit{peak performance} of both systems, we run performance-centric benchmarks similar to TPC-C and TPC-H on the cluster and the big server.
We repeated the experiments with a varying number of parallel DB clients in order to estimate a saturation point, \ie, how many parallel queries the systems can process---without entering an overload state.
In Figure \ref{figure:perf100:olap}, the OLAP benchmark results are depicted.
The x-axis shows the response time in seconds, while the y-axis illustrates the energy consumption per query.
The numbers on the individual graphs annotate the number of parallel DB clients for this curve progression.

From the figure, we can conclude that the big server handles queries generally faster than the cluster, it also exhibits better energy efficiency.
When raising the number of clients, the brawny machine takes longer to respond to queries, yet up to 340 clients, runtimes only slightly increase.
After that point, the server seems saturated and runtimes start to build up.
Consequently, energy consumption per query rises.

The cluster handles medium-sized workloads (up to 220 clients) slightly worse than the cluster.
Yet, more than 220 clients seem to overload the cluster as runtimes and energy consumption quickly increase.
When stressing the cluster with more than 340 clients, the database crashes due to shortage of main memory.

Figure \ref{figure:perf100:oltp} illustrates the results of the same experiments repeated using OLTP queries.
The results reveal that the big server is much better suited for OLTP than the cluster, as is exhibits lower query response times and also less energy consumption.
Query response times on the brawny server increase only slightly with the number of DB clients and the system does not show saturation at all.
Consequently, energy consumption per query improves continuously.

The cluster is saturated with 160 clients; when further increasing the number of parallel queries, response times start to increase faster.

Analyzing access patterns of both, OLTP and OLAP, the different performance figures are explainable:
OLAP queries read huge amounts of records, join them with (small) fact tables, and then group and aggregate the results to satisfy analytical inquiries.
Hence, the reading part of these queries can run in parallel on all partitions, speeding up the query linearly with the number of disks, CPUs, and/or nodes.
After having fetched the qualified records, the joins with the fact tables can also run concurrently.
The final grouping and aggregation steps can be pre-processed locally for each of the parallel streams and quickly aggregated into a final result.
Hence, this kind of access pattern seems to well fit both, a single, multi-core machine with lots of disks and memory but also a cluster of independent nodes, exchanging query results via network.
In \cite{CIKM1}, we have analyzed the abilities of a cluster to process that kind of workloads in greater detail .

In contrast, OLTP queries touch very little data, but update records frequently.
Since writers need to synchronize to avoid inconsistencies, lock information must be shared among all nodes involved.
A centralized machine keeping the lock table in main memory is able to synchronize transactions much quicker than a cluster, needing to exchange lock tables among nodes.
Further, OLTP query operators modifying records cannot be offloaded to other nodes.
Therefore, the query plan for transactional workloads is much more rigid than OLAP queries.

In summary, it is comprehensible, that a cluster of nodes is better suited for OLAP workloads than OLTP.

\subsection{Energy-centric benchmark}
After evaluating the peak performance of both configurations, we ran experiments representing average, real-world workloads.
Because DB servers are typically heavily underutilized, as mentioned earlier, we modified the benchmark driver to submit queries \textit{at timed intervals}.

\textbf{Workload scaling: }
In each experiment, we have spawned a number of OLTP or OLAP clients, sending queries to the database.
Each client sends a query in a specified interval.
If the query is answered within the interval, the next query is not initiated immediately, but at the start of the subsequent interval. 
If the query is not finished within the interval, the client waits for the answer until sending the next query.
In this way, each DB client generates its share of utilization.
The database has to answer queries quickly enough to satisfy the DB clients, but there is no reward for even faster query evaluation.
It is important to delay query submission of the clients, because we are not interested in maximizing throughput, but instead, want to adjust the DBMS to a given workload, using an optimal number of nodes.\footnote{Otherwise, the whole benchmark would degenerate to a simple performance-centric evaluation, which is not what we intended.}

Before each workload changes, the  cluster is manually reconfigured to best match the expected workload.
We let the benchmarks run for a short warm-up time prior to measuring energy efficiency and performance to eliminate start-up cost and to identify maximum energy savings potential.

\textbf{OLAP:}
In this experiment, we vary the number of parallel clients between 20 and 320.
As before, we are using TPC-H queries on a SF300 dataset.
To control utilization, the clients send queries at an interval of  at least 20 seconds.
Whatever comes last, the query result or the end of the interval, is the trigger for the next query. 

Figure \ref{figure:perf0-100:olap} illustrates the results for the energy-centric OLAP benchmark.
The left side depicts query response times of the brawny server and the wimpy cluster.
As expected, the centralized machine handles queries faster than the cluster, even faster than the target response time of 20 seconds per query.
Therefore, the server is idle for longer time periods, still consuming energy.

The cluster is meeting the target response times quite well, except for higher utilization, as observed earlier.
After about 220 parallel clients, query performance starts to drop and runtimes build up.

\def \figuresize {0.666}
\begin{figure*}[t]%
	\centering
    \begin{subfigure}[b]{\figuresize\columnwidth}
  		\includegraphics[width=\textwidth,page=12]{\imagedir/CIKM_OLAP_DYN}
	  \end{subfigure}%
		\hspace{0.5mm}
    \begin{subfigure}[b]{\figuresize\columnwidth}
  		\includegraphics[width=\textwidth,page=4]{\imagedir/CIKM_OLAP_DYN}
	  \end{subfigure}%
		\hspace{0.5mm}
    \begin{subfigure}[b]{\figuresize\columnwidth}
  		\includegraphics[width=\textwidth,page=8]{\imagedir/CIKM_OLAP_DYN}
	  \end{subfigure}%
		\\
		\vspace{5mm}
    \begin{subfigure}[b]{\figuresize\columnwidth}
  		\includegraphics[width=\textwidth,page=11]{\imagedir/CIKM_OLAP_DYN}
	  \end{subfigure}%
		\hspace{0.5mm}
    \begin{subfigure}[b]{\figuresize\columnwidth}
  		\includegraphics[width=\textwidth,page=3]{\imagedir/CIKM_OLAP_DYN}
	  \end{subfigure}%
		\hspace{0.5mm}
    \begin{subfigure}[b]{\figuresize\columnwidth}
  		\includegraphics[width=\textwidth,page=7]{\imagedir/CIKM_OLAP_DYN}
	  \end{subfigure}%
		\\
		\vspace{5mm}
    \begin{subfigure}[b]{\figuresize\columnwidth}
  		\includegraphics[width=\textwidth,page=10]{\imagedir/CIKM_OLAP_DYN}
	  \end{subfigure}%
		\hspace{0.5mm}
    \begin{subfigure}[b]{\figuresize\columnwidth}
  		\includegraphics[width=\textwidth,page=2]{\imagedir/CIKM_OLAP_DYN}
	  \end{subfigure}%
		\hspace{0.5mm}
    \begin{subfigure}[b]{\figuresize\columnwidth}
  		\includegraphics[width=\textwidth,page=6]{\imagedir/CIKM_OLAP_DYN}
	  \end{subfigure}%
		\\
		\vspace{5mm}
    \begin{subfigure}[b]{\figuresize\columnwidth}
  		\includegraphics[width=\textwidth,page=9]{\imagedir/CIKM_OLAP_DYN}
	  \end{subfigure}%
		\hspace{0.5mm}
    \begin{subfigure}[b]{\figuresize\columnwidth}
  		\includegraphics[width=\textwidth,page=1]{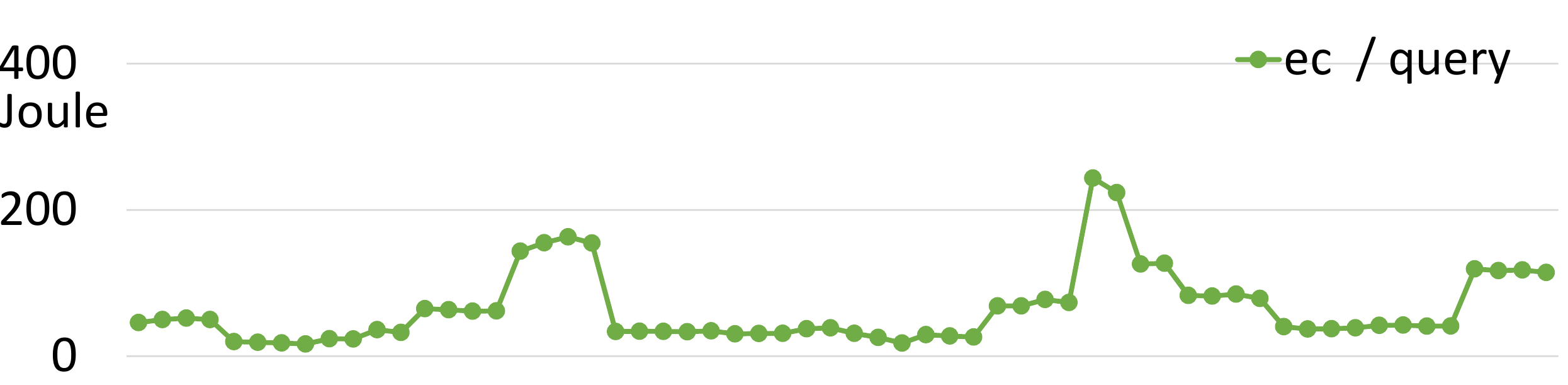}
	  \end{subfigure}%
		\hspace{0.5mm}
    \begin{subfigure}[b]{\figuresize\columnwidth}
  		\includegraphics[width=\textwidth,page=5]{\imagedir/CIKM_OLAP_DYN}
	  \end{subfigure}%
		\\
		\vspace{5mm}
    \begin{subfigure}[b]{\figuresize\columnwidth}
  		\includegraphics[width=\textwidth]{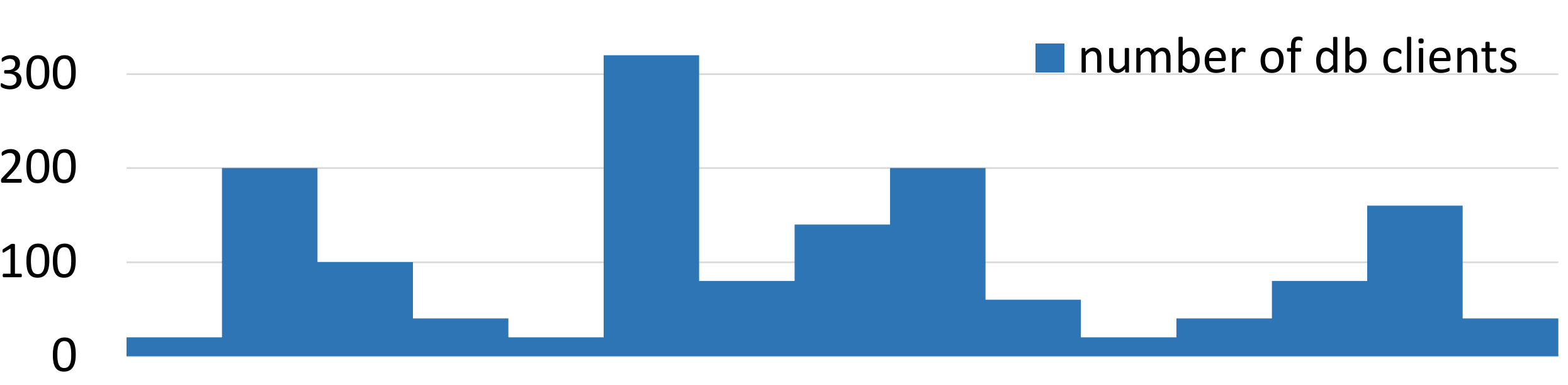}
	  	\caption{Results on the big server}%
	  \end{subfigure}%
		\hspace{0.5mm}
    \begin{subfigure}[b]{\figuresize\columnwidth}
  		\includegraphics[width=\textwidth]{\imagedir/CIKM_DYN_CLIENTS}
	  	\caption{Results without forecasting}%
	  \end{subfigure}%
		\hspace{0.5mm}
    \begin{subfigure}[b]{\figuresize\columnwidth}
  		\includegraphics[width=\textwidth]{\imagedir/CIKM_DYN_CLIENTS}
	  	\caption{Results using forecasting}%
	  \end{subfigure}%
		\caption{Dynamic OLAP workload on the big server and the cluster}
		\label{figure:dynamic:olap}
\end{figure*}

Comparing energy consumption per query of both systems, the cluster delivers far better results for \textit{average utilizations}.
Due to the cluster's scale-out and adaptation to the necessary number of nodes, its energy consumption per query stays at the same level almost the entire time, regardless of utilization.
Only at high workloads, energy consumption increases because of lengthy query runtimes.

The big server, with more or less static power consumption over the whole utilization spectrum, delivers bad energy efficiency for low and moderate workloads.
Only at high utilization, when all the processing power of the server is needed, its energy consumption per query pushes below that of the cluster.

From this experiment, we conclude that the cluster seems to be better fit for moderate OLAP workloads than the big server.

\begin{figure*}[t]%
	\centering
    \begin{subfigure}[b]{\figuresize\columnwidth}
  		\includegraphics[width=\textwidth,page=12]{\imagedir/CIKM_OLTP_DYN}
	  \end{subfigure}%
		\hspace{0.5mm}
    \begin{subfigure}[b]{\figuresize\columnwidth}
  		\includegraphics[width=\textwidth,page=4]{\imagedir/CIKM_OLTP_DYN}
	  \end{subfigure}%
		\hspace{0.5mm}
    \begin{subfigure}[b]{\figuresize\columnwidth}
  		\includegraphics[width=\textwidth,page=8]{\imagedir/CIKM_OLTP_DYN}
	  \end{subfigure}%
		\\
		\vspace{5mm}
    \begin{subfigure}[b]{\figuresize\columnwidth}
  		\includegraphics[width=\textwidth,page=11]{\imagedir/CIKM_OLTP_DYN}
	  \end{subfigure}%
		\hspace{0.5mm}
    \begin{subfigure}[b]{\figuresize\columnwidth}
  		\includegraphics[width=\textwidth,page=3]{\imagedir/CIKM_OLTP_DYN}
	  \end{subfigure}%
		\hspace{0.5mm}
    \begin{subfigure}[b]{\figuresize\columnwidth}
  		\includegraphics[width=\textwidth,page=7]{\imagedir/CIKM_OLTP_DYN}
	  \end{subfigure}%
		\\
		\vspace{5mm}
    \begin{subfigure}[b]{\figuresize\columnwidth}
  		\includegraphics[width=\textwidth,page=10]{\imagedir/CIKM_OLTP_DYN}
	  \end{subfigure}%
		\hspace{0.5mm}
    \begin{subfigure}[b]{\figuresize\columnwidth}
  		\includegraphics[width=\textwidth,page=2]{\imagedir/CIKM_OLTP_DYN}
	  \end{subfigure}%
		\hspace{0.5mm}
    \begin{subfigure}[b]{\figuresize\columnwidth}
  		\includegraphics[width=\textwidth,page=6]{\imagedir/CIKM_OLTP_DYN}
	  \end{subfigure}%
		\\
		\vspace{5mm}
    \begin{subfigure}[b]{\figuresize\columnwidth}
  		\includegraphics[width=\textwidth,page=9]{\imagedir/CIKM_OLTP_DYN}
	  \end{subfigure}%
		\hspace{0.5mm}
    \begin{subfigure}[b]{\figuresize\columnwidth}
  		\includegraphics[width=\textwidth,page=1]{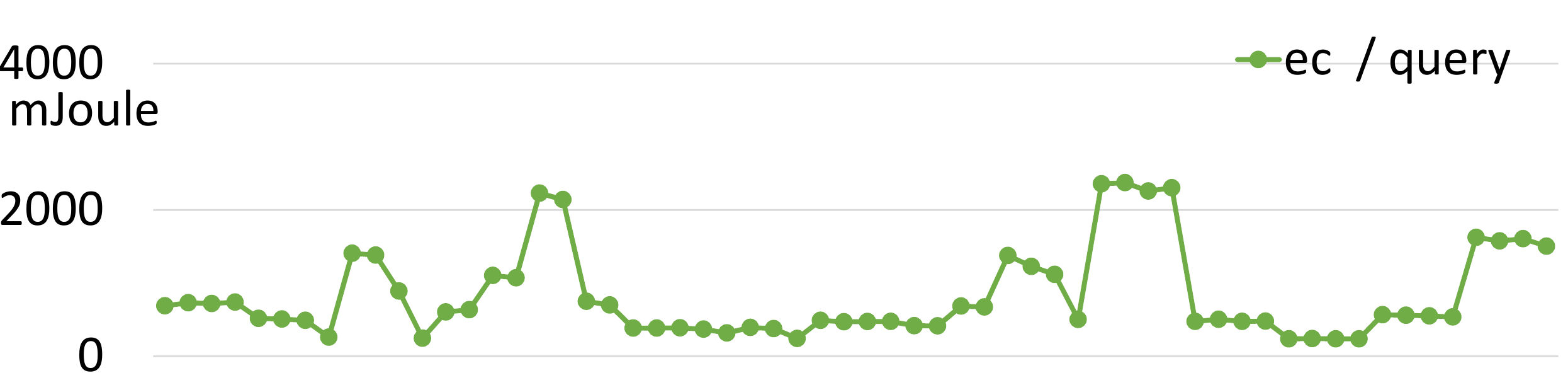}
	  \end{subfigure}%
		\hspace{0.5mm}
    \begin{subfigure}[b]{\figuresize\columnwidth}
  		\includegraphics[width=\textwidth,page=5]{\imagedir/CIKM_OLTP_DYN}
	  \end{subfigure}%
  	\\
		\vspace{5mm}
		\begin{subfigure}[b]{\figuresize\columnwidth}
  		\includegraphics[width=\textwidth]{\imagedir/CIKM_DYN_CLIENTS}
	  	\caption{Results on the big server}%
	  \end{subfigure}%
		\hspace{0.5mm}
    \begin{subfigure}[b]{\figuresize\columnwidth}
  		\includegraphics[width=\textwidth]{\imagedir/CIKM_DYN_CLIENTS}
	  	\caption{Results without forecasting}%
	  \end{subfigure}%
		\hspace{0.5mm}
    \begin{subfigure}[b]{\figuresize\columnwidth}
  		\includegraphics[width=\textwidth]{\imagedir/CIKM_DYN_CLIENTS}
	  	\caption{Results using forecasting}%
	  \end{subfigure}%
		\caption{Dynamic OLTP workload on the big server and the cluster}
		\label{figure:dynamic:oltp}
\end{figure*}

\textbf{OLTP: }
We repeated the same experiment using OLTP queries from the TPC-C benchmark.
Prior, the corresponding dataset was generated with a scale factor of 1,000.
Identical to the OLAP benchmark, we scaled the DB clients between 20 and 320.
Each client was waiting 40 ms between queries to simulate low and moderate workloads too.

Figure \ref{figure:perf0-100:oltp} plots the results of the  energy-centric OLTP benchmark run.
Whereas the big server exhibits query response times between 30 and 50 milliseconds, the cluster performs with processing times between 50 and 450 ms much worse.
Apparently, the cluster is not well suited to process update-intensive OLTP workloads.

On the other hand,  energy efficiency of the cluster is much better at low workloads.
While the big server consumes between 200 and 800 mJoule per query, the cluster only needs about 150 mJoule/query.

In conclusion, we have constituted a tradeoff between performance and energy consumption on the cluster.
By reducing the number of nodes, both power consumption and peak performance are lowered.
For moderate workloads, lower performance is tolerable and, thus, energy efficiency can be improved.

\subsection{Dynamic workloads}
As previously described, the limiting factor for dynamic repartitioning is the migration cost, \ie, the performance impact and time it takes to move data between nodes.
To estimate its impact on the cluster's elasticity, we have run experiments on a dynamically adapting cluster.
Similar to the previous tests, we are running a mix of workloads against the cluster, ranging from low utilization up to heavy workloads.
In this experiment, the cluster is given no warm-up times to adjust itself to a given task; instead, we are monitoring performance and energy consumption continuously.

Workloads change every 5 minutes, starting with a moderate workload of 20 database clients, sending OLTP or OLAP queries respectively.
The workload pattern is depicted underneath all result figures.

To quantify the importance of forecastable workloads, we have run the same workloads on the cluster twice---and for comparison once on the big server.
The first run on the cluster hits the database unexpectedly; WattDB will have to reactively adjust to the workload.
After that, the same benchmark is used again, this time informing the database of upcoming workloads (30 minutes in advance).
Hence, the database may use the information to proactively adjust.
As stated by Kramer et. al. \cite{KRAMER12}, database workloads are often repetitive and, therefore, quite easy to forecast.

In the following, results of the benchmark runs are discussed separately for OLAP and OLTP.
Results for the big server are depicted on the left side of figures \ref{figure:dynamic:olap} and \ref{figure:dynamic:oltp}.
In the middle part, the plots represent benchmark results measured with the non-forecasting cluster.
The right side illustrates results of the same experiment using forecasting.
The top-most plot in every column draws the average query response time. The target response time of 20 seconds is included to expose the load-dependent response time deviations in the various experiments.
To characterize the varying size of the cluster, the number of active nodes is visualized. 
Underneath, the course of the overall power consumption is shown for all three experiments.
The resulting average energy consumption per query is plotted in the graphs below---to contrast it to the power consumption.
The last charts in each column visualizes the workload mix (which was the same for all three experiments).

\textbf{OLAP (Big Server): }
Figure \ref{figure:dynamic:olap}, leftmost column, shows the results for TPC-H queries on the big server.
The brawny server does not exhibit transition times between workloads, since no reconfiguration is needed on this single-node system.
Query runtimes are fast, always beating the target response time.
Yet power consumption is constantly high, regardless of utilization, as already observed in earlier experiments.
Average energy consumption per query is comparably high, although query runtimes are low.
Because this benchmark is energy-centric, faster query runtimes do not lead to better results.

\textbf{OLAP (non-forecasting): }
In the middle column of Figure \ref{figure:dynamic:olap}, TPC-H results for the cluster, not using forecasting, are depicted.
The number of nodes in the cluster jitters heavily, as the system tries to adjust itself to the current workload.
Reconfiguration takes time, \eg, migrating from 2 to 4 nodes requires each of the two source nodes to ship about 100 GB of data to one of the targets, hence, it takes about 20 minutes to repartition.
Therefore, query response times in this benchmark experiment are highly fluctuating and often missing the predefined deadline.
Yet, as we have shown in the previous experiments, the cluster, in theory, should be able to handle most of the workloads within the deadline.
Due to the high additional reconfiguration overhead, the nodes are overloaded.
Therefore, query runtimes and  average energy consumption per query remain high.

\textbf{OLAP (forecasting): }
In this benchmark, we inform the cluster of workload changes present in the next 30 minutes.
Hence, instead of only reacting to workload changes, WattDB can now prepare for upcoming load.
The plots on the rightmost column of Figure \ref{figure:dynamic:olap} illustrate the results.
In comparison to the first run on the cluster, response times are generally lower and more often passing the deadline.
Because the cluster prepares for heavy workloads in advance by scaling out to more nodes, the number of nodes is also larger in average, resulting in increased power consumption.
Resulting energy consumption per query shows a mixed picture.
For low utilizations, but more nodes running to prepare for upcoming events, energy consumption is higher compared to the non-forecasting version.
On the other hand, for higher utilizations, thanks to in-advance preparations, query runtimes are lower and exhibit overall better energy efficiency.

When comparing the big server with the cluster, we can conclude that the server is more powerful and exhibits lower query response times.
On the other hand, the cluster is more energy efficient, especially during low and moderate utilization, due to its adaptation to the workload.
The cluster benefits from scale-in, when performance is not needed. This translates to a steadily varying power consumption (according to the cluster size),  whereas the server displays a more or less constant one.
For OLAP workloads, the cluster seems like an eligible alternative to a big server.

\textbf{OLTP (Big Server): }
After running OLAP benchmarks, we have repeated the same dynamic workload with OLTP clients on the TPC-C dataset.
Figure \ref{figure:dynamic:oltp} illustrates our results for energy-centric OLTP benchmark experiments.
The left column depicts those for the big server.

\textbf{OLTP (non-forecasting): }
The middle column of Figure \ref{figure:dynamic:oltp} summarizes our results for the benchmark run on a non-forecasting cluster.
Obviously, the response times shown are high. Because the cluster is forced to permanently repartition,  response times and, in turn, energy efficiency are further worsened.
Because the cluster can only react to workload changes, rebalancing starts after big workloads hit the cluster.
As discussed for the OLAP benchmark, this puts too much stress on the nodes and notably slows down query processing.
Compared to the big server, query response times are much higher for high utilizations.
Yet, power and energy consumption are lower. Therefore, the cluster delivers better energy efficiency overall---if longer query response times are deemed acceptable.

\textbf{OLTP (forecasting): }
The right-most column of Figure \ref{figure:dynamic:oltp} plots the OLTP benchmark results on the cluster using forecasting.
Compared to the previous benchmark, the average number of nodes is higher, because WattDB is preparing for workloads in advance.
As a results, query runtimes are more stable and more often pass the deadline.
However, power consumption is often higher.
Again, overall energy efficiency is characterized by a mixed picture:
Due to preparations, lower workloads have worse energy efficiency, but more intense workloads benefit from forecasting by achieving lower energy consumption per query.

\begin{figure*}[!t]%
	\centering
	\begin{subfigure}[b]{\columnwidth}
		\centering
		\includegraphics[width=0.9\textwidth]{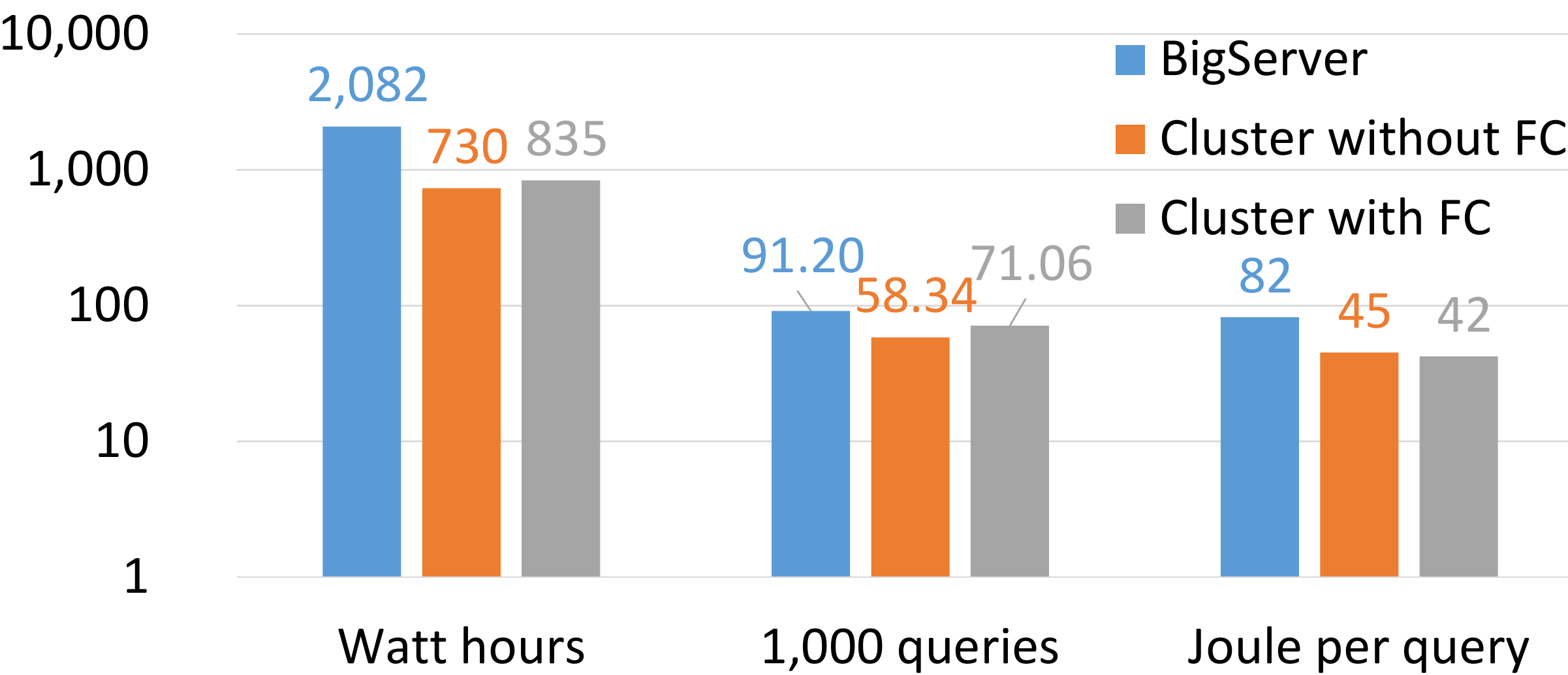}%
		\caption{OLAP workload}%
		\label{figure:dynamic:olap:summary}%
	\end{subfigure}
	\begin{subfigure}[b]{\columnwidth}
		\centering
		\includegraphics[width=0.9\textwidth]{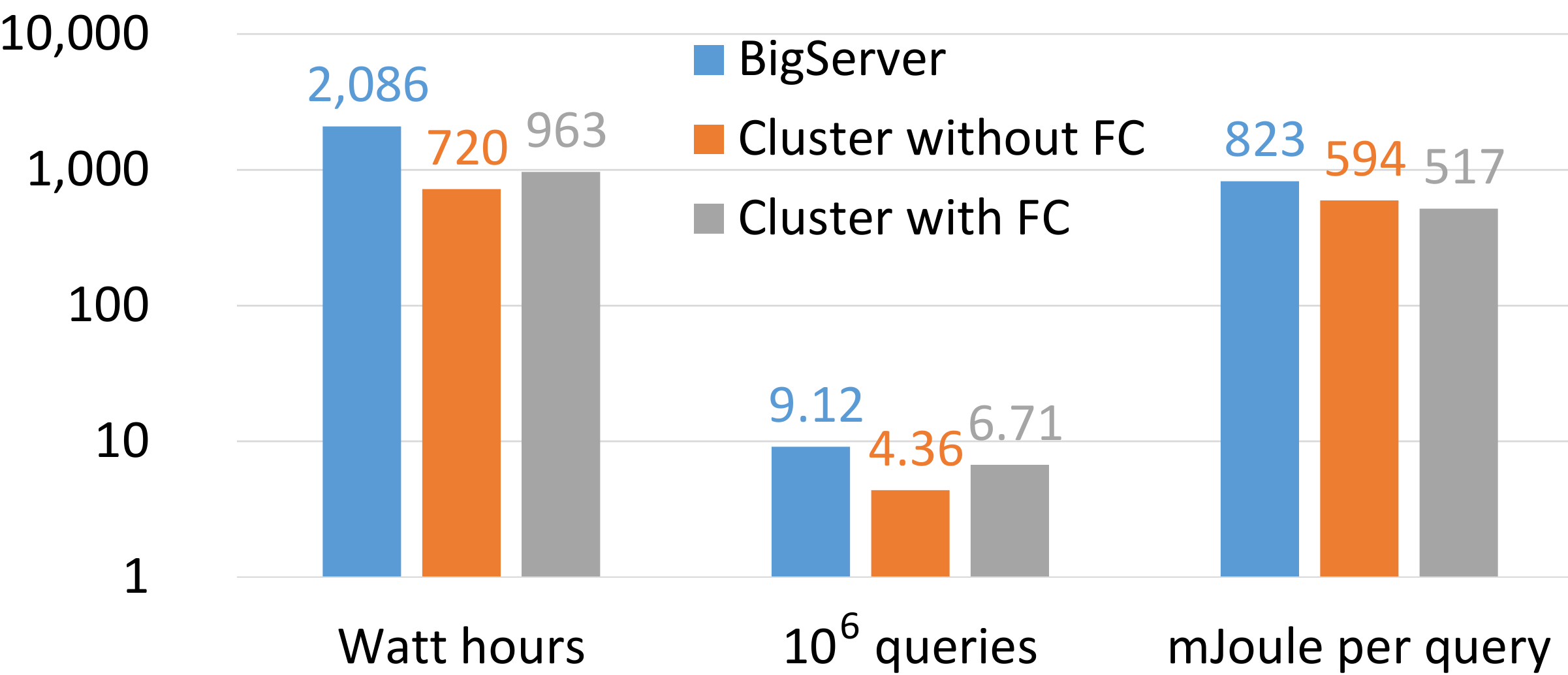}%
		\caption{OLTP workload}%
		\label{figure:dynamic:oltp:summary}%
	\end{subfigure}
		\caption{Overall energy consumption, throughput, and average energy consumption per query}%
		\label{figure:dynamic:summary}%
\end{figure*}

\textbf{Summary}
Reviewing the results from all benchmarks, we want to extract some condensed numbers to facilitate  high-level comparison and to gain a few key observations. For this reason, we have separately computed indicative numbers for the dynamic OLAP and OLTP experiments: total energy consumed (in Watt hours), overall query throughput in units of $10^{3}$ resp. $10^{6}$, and average energy consumption in Joule resp. mJoule per query. These condensed numbers are visualized in Figure \ref{figure:dynamic:summary}, where the logarithmic y-axis should be regarded.

First, the cluster is no match for the big server considering pure performance.
The centralized system does not require network communication or synchronization among nodes.
Therefore, it can deliver much better query throughput than the cluster, where queries have to be distributed, results have to be collected, and the overall execution of concurrent queries on multiple nodes needs some form of synchronization to ensure ACID properties.
All these factors lead to friction losses which slows down query processing.

Second, the cluster handles low and moderate workloads quite well, although the big server is still faster.
Yet, the cluster requires less than half of the server's power (left-most bars in the figures).
Therefore, the cluster needs less energy per query and is more energy efficient, as depicted by the right-most bars in the figures.

Third, dynamic workloads with varying utilization require preparation to adjust the number of nodes to the needs.
If workloads are predictable, the cluster exhibits better energy efficiency than the single server while delivering comparable performance.
Although, energy consumption of a forecasting cluster is higher, its query performance outweighs the additional wattage.

\section{Conclusion}
\label{section:Conclusion}
In this paper, we have examined the power-saving potential of a clustered DBMS compared to a traditional DBMS based on a single server.
An important goal of this paper was to compare performance and energy efficiency of our WattDB cluster to those of a big server.
Of course, if  peak DBMS performance is required during almost the entire operating time, a single-server approach has no alternative as our performance-centric benchmarks clearly reveal.
However, as stated in various studies \cite{BH09,TPCTC2011}, average utilization figures are far from continuous peak loads.
A large share of database or data-intensive applications runs less than an hour close to peak utilization on workdays and is resilient w.r.t. somewhat slower response times. During the remaining time, their activity level is typically in the range of 20--50\%  and often lower.
Therefore, from low- to mid-range workloads, a dynamically adjusting cluster of nodes will consume significantly less power without sacrificing too much performance.
Hence, their response time / throughput requirements could be conveniently satisfied by the performance characteristics of our cluster with much less energy use, as confirmed by Figure \ref{figure:dynamic:olap}.
Hence, the application range, where the cluster's energy efficiency largely  dominates that of a single server, has quite some practical benefit.

Especially OLAP workloads, where  lots of records need to be read and aggregated without much coordination effort, a cluster seems to be a viable alternative to a single server.
On the other hand, when processing OLTP workloads, where transactions need to synchronize continuously, a cluster suffers from high "friction losses" and is a magnitude slower than the centralized approach.
  
As shown, predictability of workloads and data elasticity are crucial for our approach.
Fortunately, typical usage patterns are predictable and a cluster can therefore prepare for upcoming workloads.
Thus, dynamically adjusting a cluster to the workload---although time-consuming---is possible.

\bibliographystyle{abbrv}
\bibliography{bibliography}

\end{document}